\documentstyle[epsfig]{aipproc}

\begin{document}
\title{Gamma Rays from Dark Matter}

\author{Raymond J. Protheroe}
\address{Department of Physics and Mathematical Physics,\\ 
The University of Adelaide, SA 5005, Australia}

\maketitle

\begin{abstract}
I give a brief review of high energy gamma-ray signatures of dark
matter.  The decay of massive $X$-particles and subsequent
hadronization have been suggested as the origin of the highest
energy cosmic rays.  Propagation over cosmological distances to
Earth (as would be the case in some topological defect origin
models for the $X$-particles) results in potentially observable
gamma-ray fluxes at GeV energies.  Massive relic particles on the
other hand, would cluster in galaxy halos, including that of our
Galaxy, and may give rise to anisotropic gamma ray and cosmic ray
signals at ultra high energies.  Future observations above 100
Gev of gamma rays due to WIMP annihilation in the halo of the
Galaxy may be used to place constraints on supersymmetry
parameter space.
\end{abstract}

\section*{Introduction}

In this brief review I shall discuss possible high energy (HE)
gamma-ray signatures of dark matter and how they may arise.  I
shall concentrate on gamma-ray signatures at high energies, and
mainly on Cold Dark Matter (CDM) as this appears to make up most
of the matter in the Universe.  Observational constraints
currently favor a so-called ``$\Lambda$CDM'' cosmological model
\cite{Primack89} in which the various contributions to the
closure parameter, $\Omega\approx 1$, are $\Omega_\Lambda \sim
0.7$ (cosmological constant), $\Omega_m \sim 0.3$ (CDM),
$\Omega_b \sim 0.045$ (baryonic), $\Omega_\nu \gtrsim 10^{-3}$
(neutrinos).  Hot dark matter would consist of neutrinos with
mass in the range 1--10 eV, and these will not be discussed
further.  CDM candidates could be broadly classified into axions,
weakly interacting massive particles (WIMPs) and supermassive
particles (wimpzillas).  Axions are predicted to explain the
absence of strong CP violation.  If they exist, axions with mass
in the range $10^{-5}$--$10^{-3}$ eV would have been produced
copiously in the early universe and could in principle make up
all the CDM, but would probably not have HE gamma-ray signatures.

Topological defects (TD) such as monopoles, cosmic strings,
monopoles connected by strings, etc., may be produced at the
post-inflation stage of the early Universe.  In the process of
their evolution the constituent superheavy fields (particles) may
be emitted through cusps of superconducting strings, during
annihilation of monopole-antimonopole pairs, etc.  These
particles, collectively called X-particles, can be superheavy
Higgs particles, gauge bosons and massive supersymmetric (SUSY)
particles.  These are generally very short-lived, and their decay
followed by a hadronization cascade could produce an observable
gamma-ray signal.  Signals of TD origin would be affected by
interactions/cascading during propagation over cosmological
distances to Earth.

There could be also superheavy quasi-stable particles with
lifetimes larger (or much larger) than the age of the Universe.
These particles could be produced by many mechanisms during the
post-inflation epoch, and survive until the present epoch.  One
interesting process is the ``gravitational production of
super-heavy particles'', in which no interaction of X-particle is
required.  Also, string theories predict the existence of other
super-heavy particles (``cryptons'') which are metastable and
could in principle form part of the CDM  (see
refs.~\cite{Kolb98,Ellis99}).  As with any other kind of CDM,
super-heavy quasi-stable X-particles would cluster in galactic
halos. The same clustering would also occur for some TD, such as
monopolonium, monopole-antimonopole pairs connected by a string,
and vortons. The gamma-ray signals from all these objects would
reach us relatively attenuated.

Perhaps the most promising WIMP CDM candidate is the lightest
SUSY particle (LSP) with mass 20--1000 GeV.  SUSY solves the
problem of the Higgs mass $m_H\to \infty$ in the Standard Model.
It is postulated that every particle has a SUSY partner with spin
${1\over 2}$ lower. It is also postulated that ``R-parity'' is
conserved, normal particles having $R=+1$, SUSY particles having
$R=-1$, and in an interaction or decay the product of $R$ being
conserved.  An important implication of R-parity is that the LSP
must be stable.  The LSP is therefore a strong candidate for CDM
and, if it exists, would be the lightest of four neutralinos:
$\tilde \chi_1$, $\tilde \chi_2$, $\tilde \chi_3$, $\tilde
\chi_4$.  Each neutralino is supposed to be a mixture of $\tilde
\gamma$, $\tilde Z$, $\tilde H^0$ and $\tilde h^0$.  Annihilation
of WIMPs could produce an observable HE gamma-ray signal.

\section*{Fragmentation functions}

If particles make up CDM they are probably WIMPS ($\chi$)
(e.g. neutralinos or heavy neutrinos) which would therefore cluster in
the halos of galaxies where they would annihilate.

TD which accumulate in galaxy halos (monopolonia,
monopole-antimonopole-pairs and vortons) could also produce a
galactic signal through the annihilation/emission and decay of short lived
``X-particles'' which would in turn decay promptly into Standard Model
states, while TD such as cosmic strings, necklaces, etc., are
extragalactic, and could produce an extragalactic signal through
the decay of short-lived X-particles.
Super-heavy quasi-stable particles
($\tau \gg t_0$) would decay similarly but these would be clustered as CDM
in galactic halos.

The annihilation (WIMP) and decay ($X$ particle) channels are then
\[
\chi \bar{\chi} \to \gamma\gamma \mbox{~or~} 
\gamma Z \mbox{~~~~~~($\gamma$-ray lines)}
\]
\[
\left\{ \begin{array}{c} \chi \bar{\chi}\\ X \end{array} \right\} \to
\left\{ \begin{array}{c} W^+W^- \\ Z^0Z^0 \\ \bar{q}q \\ e^+e^- \\
{\rm etc.} \end{array} \right\} \to \mbox{~hadrons} \to
\mbox{~($\gamma$-ray continuum)}
\]
In the second case, each decay particle or annihilation product could gives
rise to a jet of hadrons, e.g.
\[
\left\{ \begin{array}{c} \chi \bar{\chi}\\ X \end{array} \right\} \to  q\bar{q} \to \mbox{~2 jets} \to \left\{ 
\begin{array}{l}\gamma{\rm -rays}  \\ 
{\rm neutrinos}   \\ \mbox{nucleons ($\sim 5\%$)}  \\ {\rm electrons} 
\end{array} \right.
\]

\begin{figure}[htb!] %
\centerline{\epsfig{file=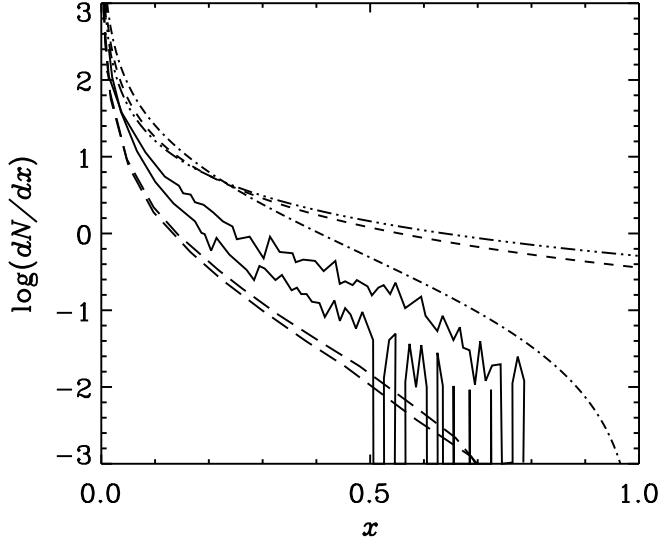,width=4.5in}}
\vspace{10pt}
\caption{Fragmentation functions for hadronization of nucleons
due to Hill\protect\cite{Hill83} (dot-dashed curve), the MLLA
approximation \protect\cite{BKV98} (short dashed curve), the
crude $x^{-1.5}$ approximation (dot-dot-dot-dashed curve), Monte
Carlo results of Birkel and Sarkar~\protect\cite{BS98} for
$m_X=10^3$GeV (upper solid curve) and $m_X=10^{11}$GeV (lower
solid curve), compared with recent results of Berezinsky and
Kachelriess~\protect\cite{BerezinskyKachelriess2000} for
$m_X=10^{12}$GeV (upper long dashed curve) and $m_X=10^{14}$GeV (lower
long dashed curve).}
\label{fig1}
\end{figure}

Energy spectra of the emerging particles, the ``fragmentation
functions'', were first calculated by Hill \cite{Hill83}.  More
recent calculations use PYTHIA/JETSET~\cite{Sjostrand2000} or
HERWIG Monte Carlo event generators to obtain the fragmentation
functions.  Each jet has energy $m_X/2$, and so one defines a
dimensionless energy for the cascade particles, $x=2E/m_X$.  The
fragmentation function for ``species a'' is then defined as
$dN_a/dx$.  A very flat spectrum of particles results, and this can
be crudely approximated by
\[
{dN_a \over dx} \propto x^{-1.5} \hspace*{6mm}
(\sim\! 32\% \; \pi^+ , \pi^0\,\mbox{and}\,  \pi^-; \sim\! 4\% N)
\] 
where the particle energies extend up to $\sim m_X/2$.  More
sophisticated treatments used the Modified Leading Logarithm
Approximation (MLLA) which is valid only for $x \ll 1$, and in more
recent QCD calculations used PYTHIA/JETSET or HERWIG Monte Carlo
event generators to obtain the fragmentation functions.  The
fragmentation functions due to Hill\cite{Hill83}, those based on the
MLLA \cite{BKV98}, and the crude $x^{-1.5}$ approximation for
large $x$ are compared in Fig.~\ref{fig1}.

Initially, the inclusion of the production of SUSY particles
\cite{BerezinskyKachelriess} was done by putting 40\% of the
cascade energy above threshold for production of SUSY particles
into LSP, thereby steepening the fragmentation functions for
normal particles at high energy.  In a recent paper Birkel and
Sarkar \cite{BS98} have shown using the HERWIG Monte Carlo that
even without inclusion of SUSY production there is a significant
dependence on $m_X$, such that for high $m_X$ the
fragmentation functions are steeper, as a direct
consequence of the well known Feynman scaling violation in
QCD~\cite{Sarkar2000}.  Fragmentation functions of Birkel and
Sarkar \cite{BS98} have been added to Fig.~\ref{fig1}, and are
well below previous QCD calculations for GUT scale $X$-particles.
Sarkar~\cite{Sarkar2000} notes, however, that the HERWIG event
generator overestimated production of nucleons by a factor
$\sim$2--3, and that new calculations by Rubin~\cite{Rubin2000}
address this issue and also include more correctly SUSY particle
production.  Very recent calculations by Berezinsky and
Kachelriess~\cite{BerezinskyKachelriess2000} (added to
Fig.~\ref{fig1}) have also used new improved treatments of SUSY
particle production, and result in only $\sim5$--12\% of the
cascade energy going into LSP.

\subsection*{Propagation over cosmological distance}

Because of the flat spectrum of particles (including gamma-rays and protons)
extending up to GUT scale energies, topological defect models
\cite{Aharonian92,Sigl96,ProtheroeStanev96} have been invoked to try to
explain the ultra high energy cosmic rays (UHE CR) for various
assumed $m_X$.  Propagation of the spectra of all particle species over
cosmological distances is necessary to predict the cosmic ray
and gamma-ray spectra expected at Earth.

For propagation of energetic particles of energy
$E$, mass $m$ and velocity $\beta c$,
through isotropic radiation the reciprocal of the mean
free path for collisions with photons is given by
\begin{equation}
x_{\rm int}(E)^{-1} = {1 \over 8 {E}^2\beta}
\int_{\varepsilon_{\rm min}}^{\infty} \, d\varepsilon
\frac{n(\varepsilon)} {\varepsilon^2} \int_{s_{\rm min}}^{s_{\rm
max}(\varepsilon,E)} ds \, (s - m^2c^4) \sigma(s),
\label{eq:mpl}
\end{equation}
where $n(\varepsilon)$ is the differential photon number density, and
$\sigma(s)$ is the relevant total cross section for a center of
momentum frame energy squared given by $s=m^2c^4 + 2 \varepsilon E(1 -
\beta\cos \theta)$ where $\theta$ is the angle between the directions
of the energetic particle and soft photon, $s_{\rm min} =
(\sum_{\rm final} m_{\rm final} c^2)^2$, $\varepsilon_{\rm min} =
(s_{\rm min}-m^2c^4 ) / [2E(1+\beta)]$, and $s_{\rm
max}(\varepsilon,E) = m^2c^4 + 2 \varepsilon E(1 + \beta)$. For
photon-photon pair production by gamma-rays, $m=0$, $\beta=1$,
$\sum_{\rm final} m_{\rm final} = 2m_e$.  For inverse Compton
scattering, $m=m_e$, $\sum_{\rm final} m_{\rm final} = m_e$.  For
Bethe-Heitler pair production, $m=m_p$, $\sum_{\rm final} m_{\rm
final} = 2m_e +m_p$, and for pion photoproduction, $m=m_p$, $\sum_{\rm
final} m_{\rm final} = m_\pi +m_p$.

\begin{figure}[htb!] %
\centerline{\epsfig{file=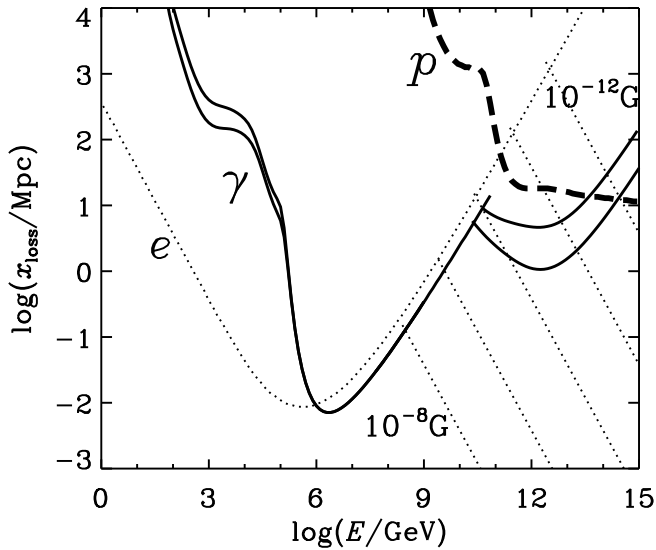,width=4.5in}}
\vspace{10pt}
\caption{Energy loss distance, i.e. mean free path divided by
inelasticity, of electrons ``$e$'', protons ``$p$'', and
gamma-rays ``$\gamma$''.  The following processes are included:
for electrons inverse Compton scattering in the cosmic microwave
background and sychrotron radiation in magnetic fields $10^{-12},
10^{-11}, \dots 10^{-8}$~G; for gamma-rays photon-photon pair
production in the cosmic infrared \protect \cite{MalkanStecker}, microwave
and radio \protect \cite{ProtheroeBiermann97} backgrounds; for protons
Bethe-Heitler pair production and pion photoproduction in the
cosmic microwave background.}
\label{fig2}
\end{figure}

The ``attenuation length'' or ``energy-loss distance'', $x_{\rm
loss}(E)$, is of greater interest and is defined by either
$x_{\rm loss}(E)=x_{\rm int}(E)/K(E)$ where $K(E)$ is the mean
inelasticity of the interaction (fraction of initial energy
lost), or $x_{\rm loss}(E)=E/(-dE/dx)$ for continuous energy loss
processes, e.g.\ synchrotron radiation.  For photon-photon pair
production $K(E)=1$.  The inelasticity is calculated using a
Monte Carlo event generator for inverse Compton interactions
\cite{Pro86,Pro90}, Bethe-Heitler pair production
\cite{ProtheroeJohnson96}, and for pion photoproduction
\cite{ProtheroeJohnson96,Muecke99}.  The radiation fields used
here are the infrared \cite{MalkanStecker}, microwave, and radio
\cite{ProtheroeBiermann97} backgrounds.  The energy loss distance
of electrons, protons, and gamma-rays is shown in
Fig.~\ref{fig2}.

As can be seen from Fig.~\ref{fig2} the Universe is very far from
being transparent above 100~GeV.  Also, protons with energies above
about $3\times 10^{11}$~GeV can travel only about 25~Mpc before losing
a substantial fraction of their energy, so that a cut-off was expected
at $\sim 10^{11}$~GeV in the cosmic ray energy spectrum if the UHE CR
originate in sources at cosmological distances, the ``GZK cut-off''
\cite{greisen,zatsepin}.  However, UHE CR have been observed above
this energy\cite{bird,akeno} and are difficult to explain by
conventional cosmic ray acceleration scenarios, making topological
defect and superheavy CDM models an attractive possibility.  

Protheroe and Meyer \cite{ProtheroeMeyer2000} have analyzed the
consequences of a recent determination of the far-infrared
background intensity.  They find that the Universe would become
nearly opaque to 20~TeV gamma rays at distances above $\sim
10$~Mpc.  Using this to correct the gamma ray flux from Markarian
501 observed by HEGRA would lead to an unacceptably high
luminosity for this source, $L_{501}\sim 10^{49}$ erg/s.  They
consider three possibilities: (i) the IR data is still
contaminated by foreground emission; (ii) the 20 TeV events are
due to Bose-Einstein condensates of lower energy photons
\cite{HarwitProtheroeBiermann99}; (iii) Lorentz Invariance (LI)
violation.  If LI violation is the explanation, then the
consequences of this are: (a) the Universe becomes transparent to
photons above 100 TeV \cite{Kifune99}; (b) the Universe also
becomes transparent to protons -- no GZK cut-off \cite{Sato99}
--- and so we should see spectra from TDs unattenuated.  In this
review, I shall adopt a conservative view, i.e.\ possibility (i), and not
consider the new far-infrared data further or the possible
consequences of it being correct, but the other possibilities
should nevertheless be borne in mind.

\subsection*{Calculating CR and $\gamma$-ray fluxes}

I give below a qualitative description of how the particle fluxes
arising from massive $X$-particle decay are calculated.
In the case of massive relic particles clustering in halos of galaxies, 
there are assumed to be two components to the flux observed at Earth:
(i) the flux due to $X$-particle decay in the halo of our galaxy,
and (ii) the flux due to $X$-particle decay elsewhere in the Universe
(mainly in halos of other galaxies).  
The fluxes can be estimated
as follows
\[
I^{h,u}_i(E)\approx{1 \over 4\pi} {n_X^{h,u} \over \tau_X} R_i^{h,u}(E)W_i(E)
\mbox{~~~~~~~~~~m$^{-2}$ s$^{-1}$ sr$^{-1}$ eV$^{-1}$}
\]
where $n_X^{h}m_x = \zeta_x \rho_{\rm CDM}^{h}$ is the density of
$X$ particles in the halo, $n_X^{u}m_x = \zeta_x \Omega_{\rm CDM}
\rho_{\rm crit}$ is the density of $X$ particles in the Universe,
$n_X$ being the number density of $X$ particles, $\zeta_x$ is
fraction of CDM in $X$ particles, and $\tau_X$ is the mean decay
time of $X$-particles.  $W^i(E)$ is energy spectrum for particle
$i$ resulting from $X$-particle decay and subsequent cascading
and is obtained from the fragmentation function.  $R^{h,u}_i(E)$
is effective size of emission region for particle $i$ produced in
our Galaxy's halo ($h$) or elsewhere in the Universe ($u$).
\[
R^{h,u}_i(E)\approx \left\{ \begin{array}{l}\mbox{size of halo
for halo intensity} \\ \mbox{attenuation length for extragalactic
intensity} \end{array} \right.
\]
In the case of $X$-particles from topological defects distributed uniformly 
throughout the Universe, only the component $I^u_i(E)$ is calculated.

\begin{figure}[htb!] 
\centerline{~~~~~~~\epsfig{file=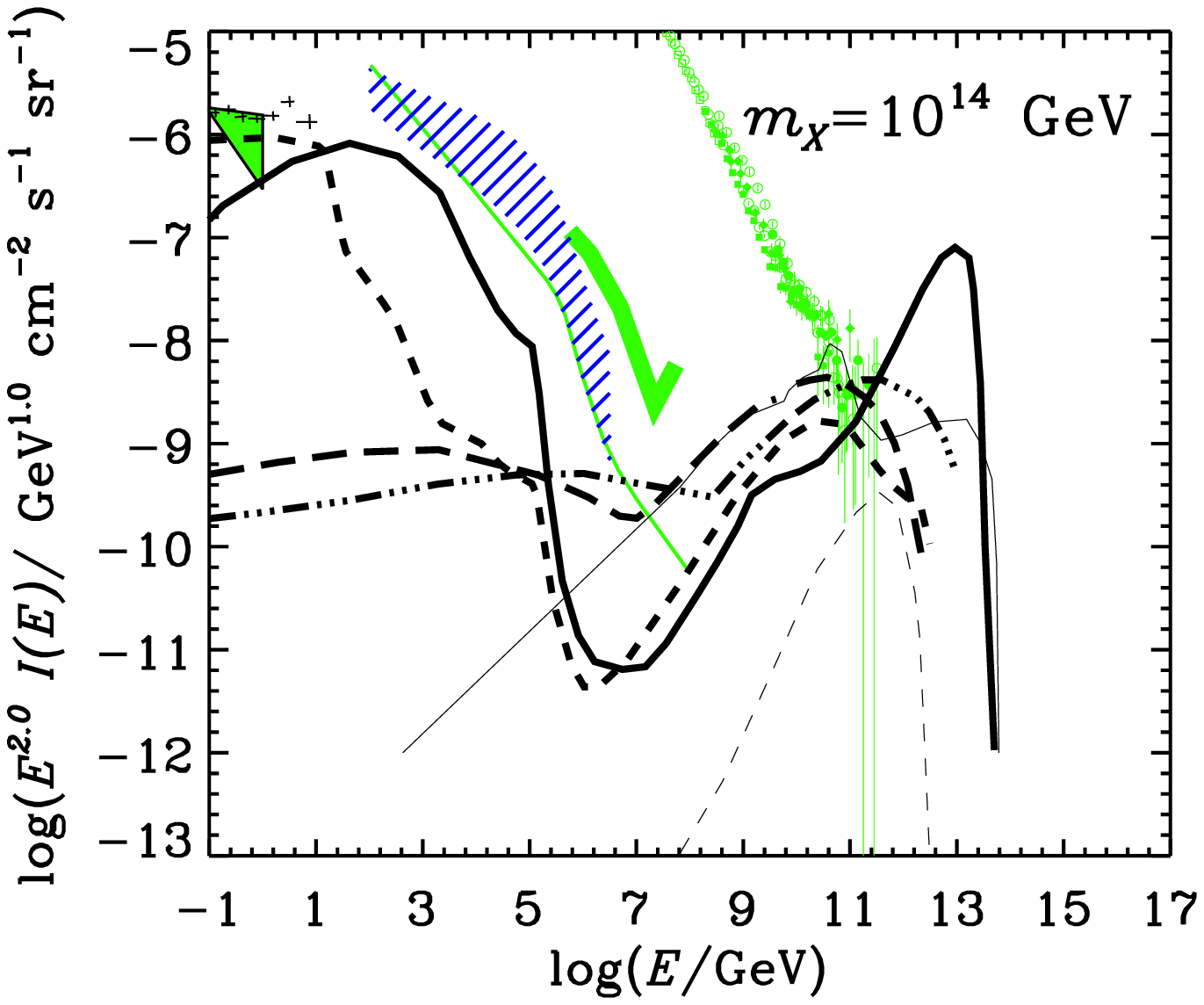,width=3.6in}\hspace*{-15mm}\epsfig{file=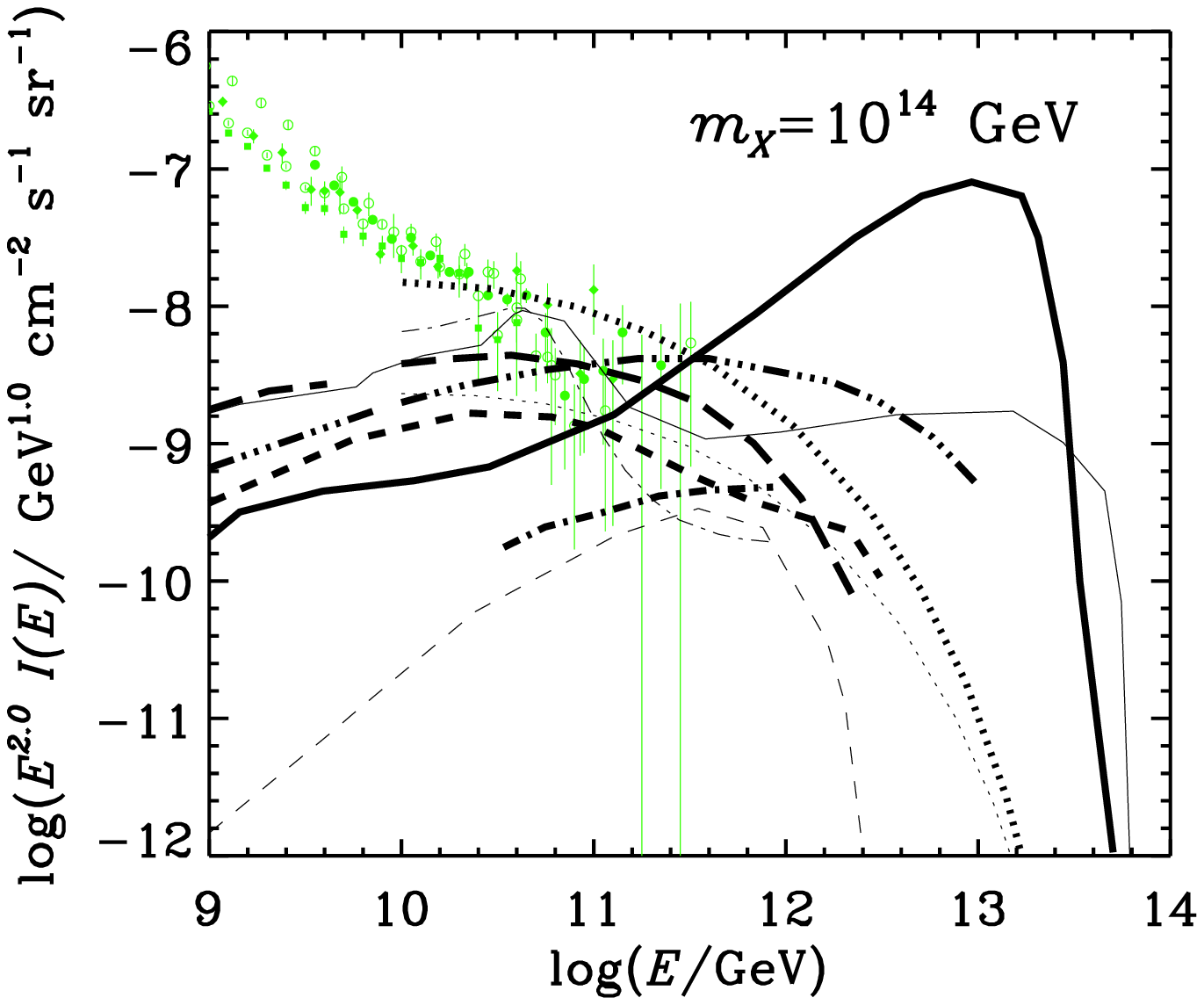,width=3.6in}}
\vspace{10pt}
\caption{Models with $m_X = 10^{14}$ GeV. Thick curves are for
gamma-rays, thin curves are for protons.  Preditions are from:
Protheroe \& Stanev \protect \cite{ProtheroeStanev96} TD (solid
curves); Sigl et al.\protect \cite{SLBY99} TD $X \to \nu\nu$ (short
dashed curves); Blasi \protect\cite{Blasi99} super-heavy relic halo population,
SUSY-QCD and $\pi\mu e$ synchrotron (gamma rays only, long dashed
curve), QCD and $\pi\mu e$ synchrotron (gamma rays only,
dot-dot-dot-dashed curve). The right panel gives more detail at UHE
energies and also includes the following results: Berezinsky et
al.\protect\cite{BBV98} super-heavy relic halo population, SUSY-QCD (dotted
curves), necklaces SUSY-QCD (dot-dashed curves).  Cosmic ray data are
taken from the survey of Gaisser and Stanev \protect\cite{GS98}.
Gamma ray data at 0.1--10~GeV are from \protect\cite{SAS2,EGRET},
gamma-ray upper limit (thick line at $10^6$--$10^8$ GeV) is from
\protect\cite{Chantell97}, cosmic ray electron inverse Compton gamma
ray prediction (shaded band at $10^2$--$3\times 10^6$ GeV) is from
ref.\protect\cite{ProtheroePorter96}, and cosmic ray $\pi^0$ gamma ray
prediction at $10^2$--$10^8$ GeV (thin curve) is from
ref.\protect\cite{IngelmanThunman}.}
\label{fig3}
\end{figure}

Sophisticated calculations take account of all cascading
processes taking place during propagation over cosmological
distances \cite{ProtheroeStanev96,Sigl96}.  In even quite low
magnetic fields synchrotron radiation by electrons dominates over
inverse Compton scattering, and Protheroe and Johnson
\cite{ProtheroeJohnson96} using a sophisticated hybrid
Monte-Carlo numerical calculation pointed out the importance of
including pair-synchrotron cascades in UHE CR propagation.
Following their approach, Protheroe \& Stanev
\cite{ProtheroeStanev96} showed that the $\gamma$-ray flux for
many TD models of UHE CR exceeded that observed at 100~MeV
energies for $B \gtrsim 10^{-9}$ G.  I show in Fig.~\ref{fig3}
their result for $m_X = 10^{14}$ GeV and $B=10^{-9}$ G (solid
curves) together with other calculations for $m_X = 10^{14}$ GeV
to be discussed later.  The flux of observable particles
($p,n,\gamma$) was normalized to the observed UHE CR flux.
Protheroe \& Stanev concluded that for higher intergalactic
magnetic fields and $m_X = 10^{14}$ GeV, or higher $m_X$ and
$B=10^{-9}$ G, the 100~MeV gamma ray flux would exceed that
observed.  Given that GUT scale $X$-particle masses were expected
(i.e. $m_X \gtrsim 10^{16}$~GeV), and extragalactic fields are
probably higher than $10^{-9}$ G it seemed in 1996 that UHE CR
could not be explained with a TD origin.

\subsection*{Recent calculations for TD and massive relic particles}

Since 1996, recent theoretical work has suggested that $m_X$ can
be significantly below the GUT mass, possibly in the range
$10^{12}$--$10^{16}$ GeV.  New fragmentation functions have been
obtained using Monte Carlo jet hadronization codes, and attempts
have been made to account for production of SUSY particles in
these cascades.  Also, new channels for $X$-particle decay have
been considered, as well as different types of TD.  Some recent
calculations have considered an extragalactic magnetic field to
be extremely low, $B \ll 10^{-9}$~G.  Interactions of UHE $\nu$
with big-bang $\nu$ have been shown to be potentially important.
Perhaps, the most interesting is the possibility of having relic
massive particles that cluster in Galaxy halos.  Having such a nearby
potential source of UHE CR means that the propagation effects for
UHE CR and gamma-rays are minimal, thereby giving a plausible
explanation to the super-GZK cosmic ray events.  

The main differences in the input to the various calculations are due
to: (1) the mass of the decaying $X$-particles, (2) the origin of the
massive particles - whether they are uniformly distributed through the
Universe (e.g. from cosmic strings, etc.), or are clustered in galaxy
halos (massive relic particles), and (3) the fragmentation functions
determined by the $X$-particle decay channel, and the particle theory
used (QCD, or SUSY-QCD).  For example, Berezinsky, Blasi \&
Vilenkin \cite{BBV98} have made calculations for ``Necklaces''
producing $X$-particles of various masses $m_X =
10^{14},10^{15},10^{16}$~GeV, and have used SUSY-QCD with 40\% of the
energy going into LSP \cite{BerezinskyKachelriess}.

In none of the models is there an absolute prediction of what the
rate of injection of primary cosmic rays is.  The predicted
fluxes are usually normalized to fit the highest energy cosmic
rays, without violating limits and observations of the diffuse
gamma-ray flux at GeV to PeV energies, and then some parameter
describing the injection rate of energy from $X$-particle decay
is determined.  One such parameter is $\zeta_Xt_0/\tau_X$, where
$\zeta_X$ is the fraction of CDM in massive $X$-particles, and
$\tau_X/t_0$ is the mean decay time of $X$-particles in units of
the Hubble time.  For example, Berezinsky et al.\cite{BKV98} have
made, predictions for several models, their preferred model
having $m_X=10^{14}$ GeV.  For $m_X=10^{13}$ GeV, using QCD and
MLLA, a halo radius $R_{\rm halo}=100$ kpc and $\Omega_{\rm
CDM}h^2=0.2$ and find $\zeta_Xt_0/\tau_X=5\times10^{-11}$.
Birkel \& Sarkar \cite{BS98} used the HERWIG QCD event generator
to get fragmentation functions, and for $m_X=10^{12}$ GeV find
$\zeta_Xt_0/\tau_X \sim 1.5\times 10^{-10}$ about 3--5 times
higher than Berezinsky et al.  They suggest that suitable
particles with $m_X\sim 10^{12}$ GeV could be Cryptons with a
decay time $\tau_X\sim 10^{20}$ yr which would give a unique
signature in the ratio of UHE neutrinos to UHE CR.

\begin{figure}[htb!] %
\centerline{~~~~~~~\epsfig{file=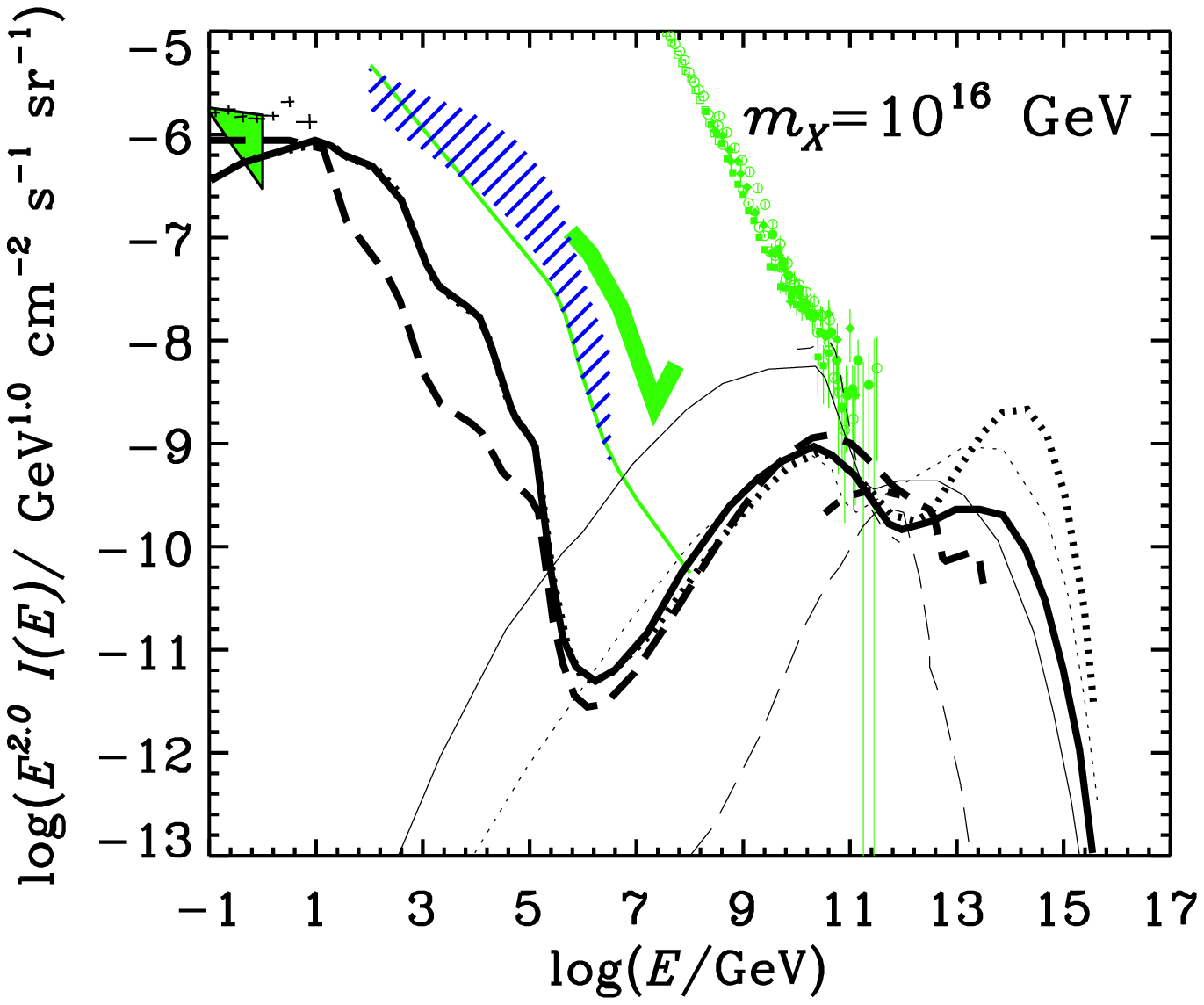,width=3.6in}\hspace*{-15mm}\epsfig{file=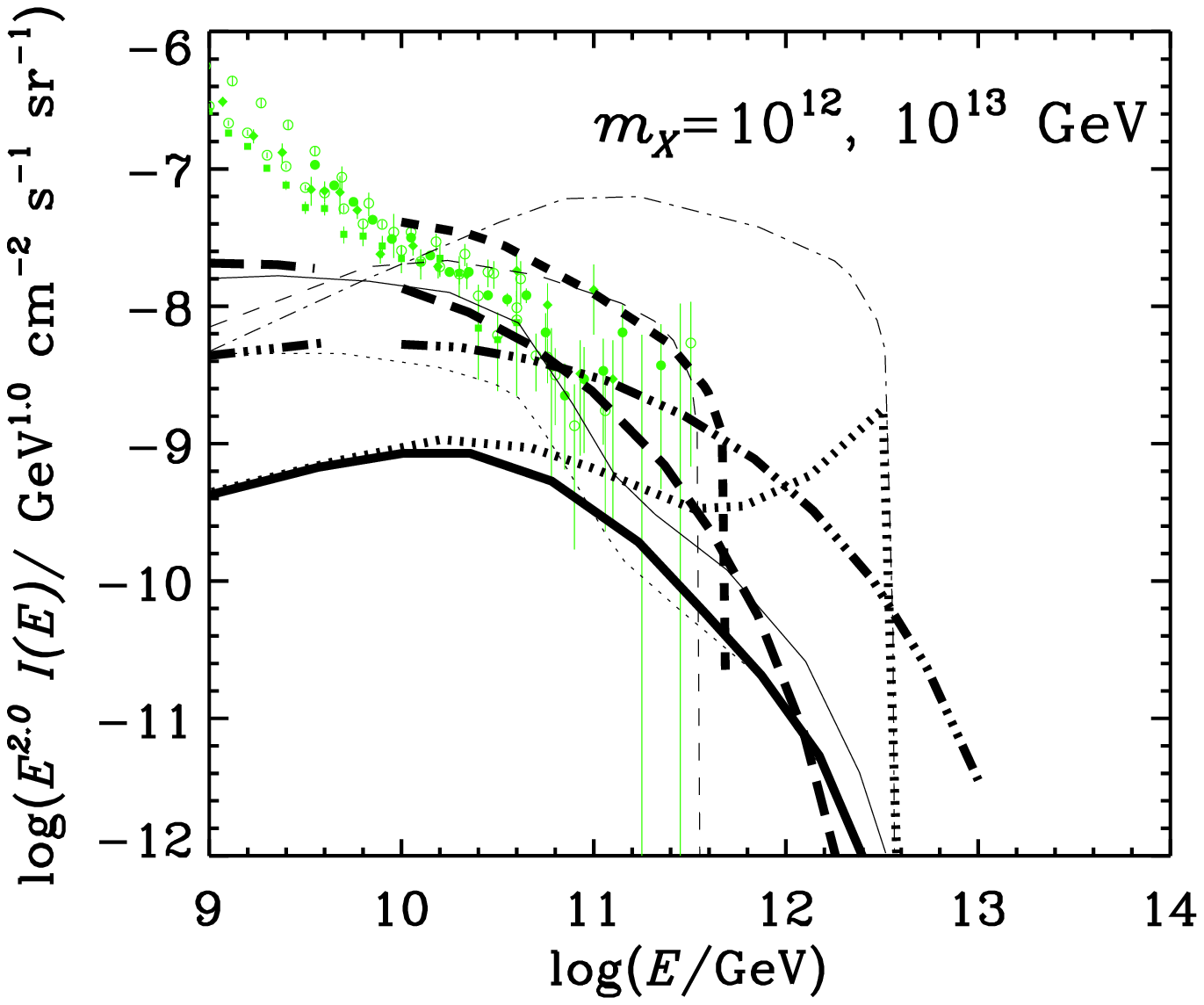,width=3.6in}}
\vspace{10pt}
\caption{(a) Models with $m_X = 10^{16}$ GeV. Thick curves are for
gamma-rays, thin curves are for protons.  Predictions are from Sigl et
al.\protect \cite{SLBY99}: TD $X \to \nu\nu$ (long dashed curves), TD
$X \to q+q$ QCD (dotted curves), TD $X \to q+q$ SUSY-QCD (solid
curves).  (b) Models with $m_X = 10^{12}$ and $10^{13}$ GeV. Thick
curves are for gamma-rays, thin curves are for protons.  Predictions
are from: Sigl et
al.\protect \cite{SLBY99} TD $m_X=10^{13}$~GeV, $X \to q+q$ QCD (solid curves), TD $m_X=10^{13}$~GeV, $X \to q+l$ QCD (dotted curves); Blasi \protect\cite{Blasi99} $m_X=10^{13}$~GeV
super-heavy relic halo population, SUSY-QCD and $\pi\mu e$ synchrotron (gamma
rays only, long dashed curve), QCD and $\pi\mu e$ synchrotron (gamma
rays only, dot-dot-dot-dashed curve); Birkel and
Sarkar\protect\cite{BS98} super-heavy relic halo population QCD
$m_X=10^{13}$~GeV (protons only, dot-dashed curve) and
$m_X=10^{12}$~GeV (short-dashed curve).  Other data and
limits are as in Fig.~\protect\ref{fig3}.}
\label{fig4}
\end{figure}

Also affecting the results, particularly for gamma-rays, are the
assumed values of the intergalactic magnetic field and the cosmic
infrared and radio background fields adopted.  It seems to me that in
many cases, unrealisticly low magnetic fields have been used
(e.g. $\le 10^{-10}$~G).  This is a problem particularly for the
models where the $X$-particles decay uniformly throughout the Universe
rather than in galaxy halos because the UHE CR must propagate over
cosmological distances to Earth, and then pair-synchrotron cascading in
realistic magnetic fields ($\ge 10^{-9}$~G) can give rise to excessive
GeV gamma-ray production at GeV energies.

In the case of decay of massive relic particles clustering in
galaxy halos being the origin of the highest energy cosmic rays,
there may well be a problem with the predicted anisotropy of
cosmic rays from our Galaxy's halo being too high
\cite{BensonSmialkowskiWolfendale99}, although this is far from
certain\cite{TancoWatson99,BerezinskyMikhailov99}.  Although
there are a variety of possible dark matter halo distributions,
the conclusions regarding anisotropy seem rather insensitive to
the model chosen.  Berezinsky, Blasi \& Vilenkin \cite{BBV98}
have made calculations for superheavy relic particles, and note
that clustering in the halo implies an anisotropy towards the
Galactic center, and an anisotropy towards Virgo.

An interesting idea due to Blasi \cite{Blasi99} concerns the case of
relic particles clustered in the halo decaying via $X \to \bar{q}q$.
He notes that since the fragmentation functions have $f_\pi \gg
f_N$, the UHE CR could be $\pi^0$ $\gamma$-rays.  In this case,
electrons from $\pi\mu e$ decay would synchrotron radiate
$\gamma$-rays which might be detectable at $E>300$ TeV.
However, it is not certain whether the highest energy cosmic rays 
can be gamma-rays (see e.g. ref.~\cite{StanevVankov97}). 

Sigl et al.\cite{SLBY99} have made a series of
calculations for the case of emission from topological defects
(no halo clustering).  An innovation in their calculations is to
include $\nu\nu$ interactions during propagation.  They consider
three decay modes: $X \to q+q$, $q+l$, or $\nu+\nu$ and
$X$-particle masses in the range $10^{13}$--$10^{16}$~GeV.
Interestingly, the $\nu\nu$ interactions during propagation
result in UHE CR protons even for the case of $X \to \nu+\nu$.

Gamma-ray signals and their associated cosmic ray fluxes for the models discussed above are shown in Figs.~\ref{fig3} and Figs.~\ref{fig4}.
Fragmentation functions for various different assumptions gives
rise to a relatively large range of model predictions.  With the
improvements in the accuracy of calculations of the fragmentation
functions~\cite{BerezinskyKachelriess2000,Rubin2000,Sarkar2000},
the current large spread in the predicted gamma ray fluxes should
hopefully decrease.

Bhattacharjee, Shafi and Stecker
\cite{BhattacharjeeShafiStecker99} point out that TDs such as
monopoles and cosmic strings associated with phase transitions in
some SUSY theories can be sources of Higgs bosons of mass $\sim
1$ TeV as well as gauge bosons of mass $\gg 1$~TeV.  These
TD-produced TeV scale Higgs may contribute significantly to the
gamma ray background above a few GeV.  The topic
of gamma-ray cascading over cosmological distances from TeV to
GeV energies in the infrared background had been discussed
earlier~\cite{ProtheroeStanev93}, and in the context of using
the observed background at GeV energies to constrain
energy injection at TeV and 
higher energies~\cite{ProtheroeStanev96,CoppiAharonian97}.

\section*{Neutralino annihilation}

The case of neutralino annihilation in CDM halos has been
discussed for several years\cite{Ellis88,Freese89}.  Assuming the CDM
to be neutralinos and considering the annihilation channels
$\chi\bar{\chi} \to \gamma \, +$ anything, one expects
a gamma-ray flux above energy $E$ of
\[
I^{\rm  SUSY}_\gamma(>E) =
{1 \over 4\pi}{\langle \sigma v\rangle N_\gamma (>\!E)
\over m_\chi^2}\int \rho_\chi^2 ds
\]
where $\sigma$ is the annihilation cross section, and $v$ is
neutralino velocity.  $\rho_\chi$ is the mass density in
neutralinos, assumed to be clustered in galaxy halos.  Because
the density is squared in the case of annihilations, the emission is
very strongly peaked towards the center of the galaxy being
observed, Baltz et al.\cite{Baltz99} suggest
looking for $\gamma$-rays from nearby galaxies.  They also point
out that if the CDM is clumped, the gamma-ray flux could be
enhanced by $\sim \times 40$ (based on estimate of the fraction of the
halo in clumps).  They assume, $\langle \sigma v\rangle N_\gamma
(>\!100\;{\rm GeV}) =10^{-25}$ cm$^{3}$ s$^{-1}$, $m_\chi=1$ TeV, 
use the Lund Monte Carlo for the fragmentation functions, and
find potentially observable fluxes above 100~GeV from within a
few arc-minutes of M87.  They note, however that a large
background from cosmic ray electrons is expected, and that an
enormous collecting area is required.

Strausz \cite{Strausz99} suggests that neutralino annihilation 
just outside the Sun by neutralinos trapped by the Sun's gravitational
potential may give a signal above 100 GeV potentially detectable 
by GLAST or MILAGRO.

Looking for gamma-rays from neutralino annihilation gives a way
of potentially exploring the SUSY parameter space.  The minimal
supersymmetric standard model (MSSM) has many free parameters,
but reasonable choices for most leave 7 remaining free.  A model,
i.e. a set of 7 parameters, is sampled by the Monte Carlo method,
and the model is rejected if it is already excluded by other
data.  For each model, the product of the annihilation cross
section with the fragmentation function for gamma-rays above energy $E$, i.e.
$\langle \sigma v\rangle N_\gamma (>\!E)$ is worked out and
plotted against $m_\chi$.  Future $\gamma$-ray observations of such
a CDM signal could in principle limit the SUSY parameter space.

Berezinsky, Bottino and Mignola~\cite{BerezinskyBottinoMignola}
noted that neutralino annihilation for the case of a power-law
galactic halo CDM density profile would produce a potentially
observable flux of gamma-rays from the galactic center (GC)
because of the cusp in $\rho^2$ expected at the GC.  More
recently, Bergstrom, Ullio and Buckley
\cite{BergstromUllioBuckley98} have studied this process in
detail, exploring the SUSY parameter space.  They calculate
$\langle \sigma v\rangle N_\gamma (>\!E)$ for a range of models
as described above, and from this obtain the expected GC $\gamma$-ray
flux for each model.  Fluxes for some of the models may be detectable
with future atmospheric Cherenkov telescopes, but for most models 
the predicted flux is below the sensitivity of any planned telescope.

\section*{GeV gamma rays from galactic halo}

The recent discovery of a diffuse galactic halo in GeV gamma-rays
by Dixon et al.\cite{Dixon98} (see
also Chary and Wright\cite{charyWright98}) has led to several
possible dark matter explanations.  For example,
Gondolo\cite{Gondolo98} suggests this may be due to annihilation
of relic WIMPS with mass $\sim 2$--4~GeV corresponding to
$\Omega_{\chi} \sim 0.1$.  Fargion et al.\cite{Fargion2000}
suggest that it could be due annihilation of heavy relic
neutrinos, $N$, with mass in the range $m_Z/2$ to $m_Z$ followed
by inverse Compton scattering of electron pairs or decay of
$\pi^0$ produced as a result of $N \to q\bar{q}$.  They conclude
that the predicted halo flux is consistent with that observed.
De Paolis et al.\cite{DePaolis99} suggest that cold H$_2$ clouds
may be clumped in dark clusters (possibly MACHOs) in the galactic
halo. Cosmic ray interactions would then produce a $\gamma$-ray
intensity at GeV energies which would be anisotropic.  For all of
these possibilities, there would be an important background due
to inverse Compton scattering from CR electrons which it might be
possible to disentangle by examining the energy spectrum of the
halo component.

\section*{Conclusion}

Most of the Matter in the Universe is CDM, and if it consists of
neutralinos, or massive relic particles they should cluster in
galaxy halos. In the case of massive relic particles, their decay
would produce UHE gamma-ray and CR signals weakly anisotropic
towards the GC, and the UHE CR spectrum would not have a GZK
cut-off.  WIMP annihilation gamma-ray signals would be strongly
anisotropic towards the GC.  Detection of such signal by GLAST
and future atmospheric Cherenkov telescopes would 
constrain SUSY models.  If CDM consists of particles associated
with Topological Defects distributed uniformly throughout the
Universe, then UHE CR are subject to the GZK cut-off.  In this
case $\gamma$-ray signals result from a pair-synchrotron cascade
in background radiation and extragalactic magnetic fields.  The
magnetic fields used in some cascade calculations may have
usually been unrealisticly low, and the infrared and radio
radiation fields are subject to uncertainties.  Absolute fluxes
for TD models are not predictable, but detection of UHE gamma ray
signals can be used to constrain models.  Currently, predictions
for the case of massive relic particles and TD models suffer from
uncertainties in the fragmentation functions as illustrated by
the spread in Fig.~\ref{fig1}, but work is underway to improve
this~\cite{BerezinskyKachelriess2000,Sarkar2000}.

\subsection*{Acknowledgments}

I thank Venya Berezinsky and Subir Sarkar for reading the original manuscript
and making helpful suggestions.

\end{document}